# Nature-nurture interplay in educational attainment

Dilnoza Muslimova[1,4,*], Hans van Kippersluis[1,4], Cornelius A. Rietveld[1,2,4], Stephanie von Hinke[1,3,4], S. Fleur W. Meddens[1,5]

[1] Erasmus School of Economics, Erasmus University Rotterdam, The Netherlands
[2] Erasmus University Rotterdam Institute for Behavior and Biology, The Netherlands
[3] School of Economics, University of Bristol, United Kingdom
[4] Tinbergen Institute, The Netherlands
[5] Statistics Netherlands

## Abstract

This paper shows how nature (i.e., one's genetic endowments) and nurture (i.e., one's environment) interact in producing educational attainment. Genetic endowments are measured using a polygenic score for educational attainment, while we use birth order as an important environmental determinant of educational attainment. Since genetic endowments are randomly assigned within-families and orthogonal to one's birth order, our family fixed effects approach exploits exogenous variation in genetic endowments as well as environments. We find that those with higher genetic endowments benefit disproportionally more from being firstborn compared to those with lower genetic endowments.

**Keywords:** Gene-environment interplay, birth order, polygenic score/index
**JEL codes:** D19, I24, J24
**Word count:** 15830

**Acknowledgments:** *Corresponding author: muslimova@ese.eur.nl. Co-authors' emails: hvankippersluis@ese.eur.nl, nrietveld@ese.eur.nl, s.vonhinke@bristol.ac.uk, meddens@ese.eur.nl. This research has been conducted using the UK Biobank Resource (Application Number 41382). UK Biobank has received ethical approval from the National Health Service Northwest Centre for Research Ethics Committee (11/NW/0382) and has obtained informed consent from its participants. The authors gratefully acknowledge funding from NORFACE through the Dynamic of Inequality across the Life Course (DIAL) programme (462-16-100). Research reported in this publication was also supported by the National Institute on Aging of the National Institutes of Health (R56AG058726). C.A.R. and S.v.H. gratefully acknowledge funding from the European Research Council (GEPSI 946647; DONNI 851725). S.F.W.M. gratefully acknowledges funding from the European Union's Horizon 2020 research and innovation program under the Marie Skłodowska-Curie grant agreement (GENIO 101019584). We thank Aysu Okbay and employees and participants of the 23andMe, Inc. for sharing GWAS summary statistics for educational attainment. We also thank Pietro Biroli, Gabriella Conti, Ben Domingue, Monique de Haan, Paul Hufe, Owen O'Donnell, Kjell Salvanes, Dinand Webbink, Herman van de Werfhorst, and seminar participants at the NORFACE DIAL workshop on intergenerational mobility and human capital accumulation, the Vienna University of Economics and Business, DIAL Mid-term Conference (University of Turku), RSF Summer Institute in Social Science Genomics, Health Economics Symposium (Erasmus University Rotterdam), the Applied Microeconomics and Gender Workshop (University of Alicante), and the Dutch Economists Day for providing constructive comments. This work made use of the Dutch national e-infrastructure with the support of the SURF Cooperative (EINF-1107).

# I. Introduction

It is commonly hypothesized that important life outcomes such as educational attainment are influenced by a complex interplay between nature (i.e., genetic variation) and nurture (i.e., environments; Rutter, Moffitt, and Caspi, 2006; Heckman, 2007). Empirically estimating the separate contributions of genetic endowments and environments and their possible interaction, however, is complicated by the endogenous nature of both. Indeed, environmental characteristics are partially heritable and typically cluster together: e.g., higher educated parents tend to have higher incomes. Similarly, genetic variation is only random conditional on parental genotypes (e.g., Lawlor, Harbord, Sterne, Timpson, & Davey Smith, 2008; Davey Smith & Ebrahim, 2003). Not controlling for the latter implies that the 'genetic-effect' may in fact reflect 'genetic nurture' – that is, the parental genotype can shape the environment in which children grow up, thereby producing a spurious association between the child's genetic endowments and their outcomes (e.g., Belsky et al., 2018; Kong et al., 2018).

This study exploits exogenous variation in both genetic endowments for education and the family environment to analyze the importance of causal gene-environment interactions for educational attainment. Genetic endowments are measured using a so-called "polygenic score" (PGS), also referred to as a "polygenic index" (Becker, Burik, et al., 2021). A PGS is a highly predictive index that is constructed as the sum of all measured genetic variants, weighted by the strength of their relationship with educational attainment as estimated in an independent sample (Dudbridge, 2013; Lee et al., 2018). The PGS for educational attainment can currently explain up to ~15% of its variation (Okbay et al., 2022). Our measure of the environment is an individual's birth order, which is consistently negatively correlated with educational attainment (in developed countries; see e.g. Bagger et al., 2021; Behrman et al., 1986; Black, Devereux, & Salvanes, 2005; Booth & Kee, 2009; De Haan, 2010; De Haan,

Plug, & Rosero, 2014; Kantarevic et al., 2006).

We overcome endogeneity issues by exploiting within-family variation in both birth order and genetic endowments. Siblings' birth order is randomly determined within families (e.g., Damian and Roberts, 2015), and genetic variants are randomly assigned across siblings ("Mendel's Law of Independent Assortment"). Hence, within-family, genetic variants are unrelated to birth order by construction.[1] With exogenous variation in both genes and environments, this provides a compelling context in which we can fundamentally improve our understanding of the nature-nurture interplay in shaping educational attainment.

Our empirical analysis exploits data from 14,850 full siblings from the UK Biobank, a population-based sample from the United Kingdom (Fry et al., 2017). To measure participants' genetic endowments, we construct polygenic scores for educational attainment based on the results from our own tailor-made genome-wide association study (GWAS) that uses the UK Biobank sample but excludes all siblings and their relatives.[2] We adopt a family fixed effects approach to exploit within-family variation in genetic endowments and birth order to study the genetic (G), environment (E) and interaction (G×E) effects on educational attainment, and we apply Obviously-Related Instrumental Variables estimation (ORIV; Gillen, Snowberg, & Yariv, 2019) to overcome random measurement error in the polygenic score. We mainly focus on firstborns versus laterborns, since the literature suggests that birth order effects are particularly salient at this margin (e.g., Breining et al., 2020).

We confirm earlier findings that firstborns have a higher level of education than laterborns, and that one's genetic endowments are a strong predictor of educational attainment. We also confirm that genetic endowments do not differ systematically across birth order within a family. This finding corroborates that birth order effects are due to environmental influences (see also Isungset, Freese, Andreassen, & Lyngstad, 2022). Our main finding is that birth order and genetic endowments interact: those with a higher polygenic score benefit disproportionally more from being firstborn compared to those with a lower polygenic score.

[1] A systematic relationship between birth order and genetic endowments could arise when parents base their fertility decisions on the observed genetic endowments of their offspring, i.e., a stopping rule depending on the "quality" of children (e.g., Eirnæs & Pörtner, 2004). We find no such evidence in our sample (see Section 4.5).

[2] See Appendix A for definitions and explanations of the genetic terms used here.

Our results are quantitatively meaningful. We find that one's genetic endowments amplify the effect of birth order: a firstborn sibling with a polygenic score two standard deviations above the mean on average enjoys a full year of schooling extra compared to a laterborn sibling with a similar polygenic score. Handy & Shester (2022) estimate that the rise in the fraction of laterborn children was responsible for 20-35% of the stagnation in college completion among US baby-boom cohorts born between 1946 and 1974. Hence, the random inheritance of genetic endowments across siblings alongside the arbitrary order of birth have important monetary and non-monetary consequences across siblings (e.g., Heckman, Humphries, & Veramendi, 2018) as well as their offspring (e.g., Havari & Savegnago, 2022; Mogstad & Torsvik, 2022).

Our findings provide one of the first pieces of causal evidence of how genetic variation (here measured by the polygenic score for education) and the environment (here measured by birth order) jointly shape and interact in producing important life outcomes such as educational attainment. While this finding was long anticipated (e.g., Heckman, 2007; Rutter et al., 2006), finding credible and independent sources of variation in genes and environments is rare given how tightly genetic and environmental influences are entangled (e.g., Koellinger & Harden, 2018). Showing evidence of an interaction between genetic variation and environments is therefore a step forward in our fundamental understanding of how nature and nurture jointly shape human capital, while at the same time providing an antidote against arguments of genetic or environmental determinism.

Our paper speaks to two main literatures. First, we contribute to an emerging literature on gene-environment interactions (G×E), which addresses how the environment moderates the effect of genetic variants, and vice versa. Previous studies have typically examined interactions between endogenous (i.e., between-family) polygenic scores and endogenous environments such as socio-economic status (see e.g., Barth, Papageorge, & Thom, 2020; Bierut, Biroli, Galama, & Thom, 2018; Ronda et al., 2022), childhood trauma (e.g., Mullins et al., 2016; Peyrot et al., 2014), or a partner's death (e.g., Domingue, Liu, Okbay, & Belsky, 2017). Interpretation of these findings is not straightforward, because individuals with certain genetic predispositions may self-select into different environments (known as gene-environment correlation, or *rGE*; see Jencks, 1980; Schmitz & Conley, 2017a). In these analyses, therefore, the 'environmental effect' could be reflecting the effect of one's genotype through *rGE*, and the 'genetic effect' could be reflecting the rearing environment shaped by the parental genotype (i.e., genetic nurture).

A handful of studies use exogenous variation in environments to study G×E in educational attainment. For example, Conley & Rauscher (2013) analyze how random differences in the prenatal environment alter the genetic effects on education; Schmitz & Conley (2017b) use the Vietnam War conscription as a natural experiment; Barcellos, Carvalho, & Turley (2021) use a UK compulsory schooling reform; von Hinke & Sorensen (2022) explore an unanticipated peak in pollution; Muslimova (2022) uses legal access to contraceptive pill in the UK; Van den Berg, Von Hinke, & Wang (2022) use the unexpected temporary post-war derationing of sugar confectionery, and Van den Berg, Von Hinke, & Vitt (2022) exploit the introduction of the measles vaccine to study G×E effects on individuals' educational attainment.[3] We push this literature further by not only considering exogenous variation in the environment, but also in genetic endowments by exploiting within-family variation in polygenic scores.

A second strand of literature that we speak to is the literature on birth order effects. This literature consistently finds that in developed countries, laterborn children have lower educational attainment. Birth order effects have also been found for other outcomes, though sometimes with mixed results, such as early life cognitive skills and intelligence (Black, Devereux, & Salvanesz, 2011; Fenson et al., 1994; Keller, Groesch and Trob, 2015), health (Black, Devereux, & Salvanes, 2016; Pruckner et al., 2019), personality and leadership skills (Black, Grönqvist, & Öckert, 2018), and delinquency (Breining et al., 2020). Interestingly, earlier literature found only very limited heterogeneity in the birth order effects across sociodemographic factors, with only some tentative evidence that birth order effects are stronger for children of higher educated mothers (e.g., Black, Devereux, & Salvanes, 2005; Booth & Kee, 2009). We contribute to this literature by studying heterogeneity in the birth order effect on educational attainment with respect to genetic endowments. The interaction between birth order and genetic endowments is not merely an important source of previously hidden heterogeneity in the treatment effect, but one that carries over to the next generation, potentially exacerbating intergenerational inequalities (Barclay, Lyngstad, & Conley, 2021; Havari & Savegnago, 2022).

## II. Empirical Strategy

We analyze gene-environment interactions (G×E) between genetic endowments

[3] Studies with other outcomes that exploit exogenous environments include, e.g., Barban, Cao, & Francesconi (2021), Barcellos, Carvalho, & Turley (2018), Dias Pereira, Rietveld, & van Kippersluis (2022), Schmitz & Conley, (2016). For a recent review of the G×E literature in economics and social science, see Dias Pereira et al. (2022).

for educational attainment and birth order. Following Todd & Wolpin (2003) and Cunha & Heckman (2008), we specify a linear production function, where years of completed education for individual *i* of family *j* (denoted by $Y_{ij}$), is a function of initial endowments, the environment shaped by parents, and unobserved parental characteristics. Empirically, we measure initial endowments by the polygenic score for educational attainment (denoted by $G_{ij}$)[4], the environment is measured by an indicator for being firstborn ($E_{ij}$), and parental characteristics are subsumed into the family fixed effect ($\delta_j$). We then allow for an interaction term between $G_{ij}$ and $E_{ij}$, leading to the following specification:

$$Y_{ij} = \alpha_1 + \alpha_2 G_{ij} + \alpha_3 E_{ij} + \alpha_4 \left(G_{ij} \times E_{ij}\right) + \alpha_5 X_{ij} + \delta_j + \xi_{ij}, \quad (3)$$

where $X_{ij}$ is a set of individual level controls, including gender, month, and year of birth dummies (Black et al., 2005; Handy & Shester, 2022). It also includes the vector of the first 40 principal components (PCs) of the genetic relatedness matrix.[5] Finally, $\xi_{ij}$ is the error term; we use heteroskedasticity-robust standard errors, clustered at the family level. The parameter $\alpha_2$ captures the association between the polygenic score for education and years of schooling, whilst $\alpha_3$ estimates the average advantage in years of schooling for firstborn children compared with their laterborn siblings. $\alpha_4$ shows the extent to which the polygenic score and being first born complement each other's effect on education and is therefore informative about the existence of putative G×E effects.[6] Following Black et al. (2005, 2011), Heiland (2009), Lehmann, Nuevo-

[4] For ease of interpretation, we standardize $G_{ij}$ to have mean zero and standard deviation one. See Section 3 and Appendix A for more information about its construction.

[5] Genetic principal components (PCs) can be used to control for subtle forms of population stratification (i.e., correlations between allele frequencies and environmental factors across subpopulations in the sample) in a between-family (population-level) analysis (Price et al., 2006; Rietveld et al., 2014). Although not strictly necessary to include in the within-family analysis due to the inclusion of family fixed effects, we keep the PCs in all specifications to facilitate a clean comparison between the between-family and within-family results. Including them does not affect our estimates.

[6] In Section 4.5, we study the interaction effect using more flexible approaches than the linear interaction presented here, and we explore the issue that an interaction term in a family fixed

Chiquero, & Vidal-Fernandez (2018), and Amin et al. (2021), we compare specifications with and without family fixed effects. The inclusion of family fixed effects ensures that variation in the polygenic score and birth order is random, ensuring polygenic score and birth order are orthogonal to each other; we verify this empirically in Section 4.3.[7]

Our polygenic score incorporates two sources of measurement error. First, since the GWAS underlying the construction of a polygenic score is based on a finite sample (see Section 3), our estimated polygenic score that is based on the GWAS estimates is a noisy proxy for the true (latent) polygenic score (Van Kippersluis et al., 2022). Second, GWASs (the estimates of which we use to construct the polygenic score) typically do not control for parental genotypes (Howe et al., 2022). Trejo & Domingue (2019) show that not controlling for parental genotypes causes the polygenic scores to be measured with error. Both sources of measurement error lead to an attenuation bias in the coefficient of the polygenic score in within-family analyses.[8] While we cannot solve the attenuation bias arising from the omission of parental genotype in the GWAS, we follow DiPrete, Burik, & Koellinger (2018) and Van Kippersluis et al. (2022) in effects model may be partially identified from between-family variation (Shaver, 2019; Giesselmann and Schmidt-Catran, 2022).

[7] Another way to obtain a causal interpretation of the coefficient on the polygenic score is to include the *parental* polygenic scores as control variables. We do not follow this strategy here since the purpose of our family fixed effect-specification is to ensure random variation in the polygenic score *as well as* birth order. Including the parental polygenic scores instead of family fixed effects does not deal with the endogeneity of birth order (see also Appendix H).

[8] The reason why classical measurement error as a result of finite-sample bias leads to attenuation bias is well known. It is more subtle why the exclusion of parental genotype in the discovery GWAS leads to an attenuation bias in within-family studies. The reason is that a polygenic score constructed on basis of a GWAS that did not control for parental genotype will reflect both direct genetic effects arising from the individual's genotype as well as indirect genetic effects arising from the omitted parental genotypes. The latter effects are known as 'genetic nurture' (e.g., Kong et al., 2018). When applied within families, the differences in the polygenic score arising from parental genotype are spurious since parental genotype is the same across siblings. Hence, part of the differences across siblings in the polygenic score reflect genetic nurture (i.e., indirect genetic effects) and can be considered measurement error attenuating the resulting within-family estimates (Trejo & Domingue, 2019).

applying Instrumental Variables (IV) to tackle the classic measurement error problem. More specifically, we split our discovery GWAS sample into two equal halves and construct two polygenic scores based on the two discovery samples. To ensure that these subsamples are unrelated, we randomly select only one individual from each cousin cluster. Even though the two polygenic scores individually have lower predictive power, the measurement error in the two is plausibly orthogonal and so they can be used as instrumental variables for each other. Using these two polygenic scores, we apply Obviously-Related Instrumental Variables (ORIV; Gillen et al., 2019); see Appendix C for more details.

## III. Data

We use data from the UK Biobank, a population-based cohort with approximately 500,000 individuals aged between 40-69 at the time of interview in 2006-2010 and living within a radius of 35 km from one of the 22 assessment centers in England, Wales, and Scotland (Fry et al., 2017). It contains survey data, biomarker and DNA samples, physical measurements, and linkage to inpatient registers and death records (Sudlow et al., 2015). Because participation in the UK Biobank is voluntary, it is not a representative sample of the UK population (see Fry et al. (2017) for a detailed analysis).

We apply the following sample selection criteria. The original data include 502,498 consented individuals. We follow the literature and remove those of non-European descent (92,892 observations), as well as twins and multiple births (9,310 observations), and individuals with missing or conflicting information regarding the number of siblings and/or family size (3,801 observations). In doing so, we arrive at a sample of 396,494 individuals. We further restrict this sample to individuals with at least one sibling[9] in the UK Biobank and without missing values on any of the variables included in our analysis (i.e., years of education, birth order, family size, year and month of birth, principal components of the genetic relatedness matrix, gender, and the polygenic score for education). Since the UK Biobank did not specifically target families, this leads to a final sample size of *N* = 14,850 individuals within 7,281 full-sibling clusters.[10]

***Table 1.*** *Descriptive statistics analysis sample (N = 14,850).*

| Variable | Mean | S.D. | Min. | Max. |
|---|---|---|---|---|
| Years of education | 15.058 | 4.951 | 7.000 | 20.000 |
| Firstborn (1/0) | 39.4% | | | |
| PGS for years of education | 0.000 | 1.000 | -4.049 | 3.912 |
| Birth order | 1.913 | 0.998 | 1.000 | 5.000 |
| Second born | 41.50% | | | |
| Third born | 11.18% | | | |
| Fourth born | 4.33% | | | |
| Fifth- or laterborn | 3.60% | | | |
| Family size | 2.987 | 1.530 | 2.000 | 14.000 |
| Last child (1/0) | 36.9% | | | |
| Male (1/0) | 42.5% | | | |

*Notes:* PGS = Polygenic score; S.D. = Standard deviation; Min. = Minimum; Max. = Maximum.

We follow the literature (see e.g., Lee et al., 2018; Okbay et al., 2016, 2022; Rietveld et al., 2013) and convert individuals' qualifications to equivalent years of education using the International Standard Classification of Education (ISCED).[11] The average years of education for the sibling sample is 15 (see Table

[9] Siblings are identified based on the genetic data; there is no direct information about sibling status in the UK Biobank. We also use the genetic data to identify other relatives up to the third degree. See Appendix B for the full procedure followed to identify siblings and relatives.
[10] The descriptive statistics in our analysis sample are very close to those in the full UK Biobank sample.
[11] Years of education ranges from 7 to 20, where College or University degree is equivalent to 20 years, National Vocational Qualification (NVQ), Higher National Diploma (HND), or Higher National Certificate (HNC) to 19 years, other professional qualifications to 15 years, having an A or AS levels or similar to 13 years, O levels, (General) Certificate of Secondary Education ((G)CSE) to 10 years, and if none of the above to the lowest level of 7 years.

1). We construct individuals' birth order based on their response to a question of how many older siblings they have.[12] If a respondent reports zero older siblings, the birth order is set to one. For individuals with missing information on the number of older siblings, we determine birth order based on family size and birth year of the individual and his/her siblings if all of them are present in the UK Biobank. This adds information on birth order for 1,752 siblings in our analysis sample. Table 1 shows that 39.4% of our sample are firstborns, with an average birth order of 1.91 (where we have censored birth order at 5 for the 245 respondents with birth order beyond 5). Around 37% of our sample is lastborn, and the average family size is 3 (i.e., the average number of siblings is 2).

Our measure of genetic endowment for education is the polygenic score for education. A polygenic score is a weighted sum of genetic variants called Single Nucleotide Polymorphisms (SNPs, see Appendix A for details). The SNP weights are determined by the association between a SNP and years of education (Dudbridge, 2013) in an independent (discovery) sample:

$$PGS_i = \sum_{k=1}^{K} \beta_k x_{ik}, \tag{4}$$

where $PGS_i$ is the value for the polygenic score for individual *i*, $\beta_k$ is the regression coefficient of SNP *k* ($k = 1, \dots, K$) from our own tailor-made GWAS (see Appendix A), and $x_{ik}$ is the genotype of individual *i* for SNP *k* (coded as 0, 1 or 2, indicating the number of "effect" alleles). We use the LDpred software (Vilhjálmsson et al., 2015) to correct for the correlation structure across SNPs. The polygenic scores are standardized within the sibling sample to have mean 0 and standard deviation 1.

[12] Birth order is based on the number of older siblings, reported by 186,229 UKB participants. Hence, the construction of birth order variable does not require us to observe all siblings in the UK Biobank.

# IV. Results

## 4.1. Predictive power of the polygenic score for educational attainment

Figure 1 shows that our polygenic score for years of education is approximately normally distributed. The scatterplot divides the polygenic score distribution in 200 bins; the dots represent the average years of education for each bin. The line through the dots is obtained from a local polynomial regression of years of education on our polygenic score. In line with the literature, the polygenic score is positively correlated with years of education ($r = 0.23$, $p < 0.001$). The average difference between those two standard deviations below the mean of the polygenic score and those two standard deviations above the mean is almost 4 years, highlighting the substantial predictive power of the polygenic score. Furthermore, Figure 1 suggests the relationship is approximately linear.

Table C.1 in Appendix C shows that the incremental $R^2$ of the polygenic score (i.e., the additional variance explained by the polygenic score after controlling for gender, month, and year of birth, and the first 40 principal components of the genetic relatedness matrix) is 5.4% in the analysis without family fixed effects (i.e., 0.113-0.059=0.054). In the specifications with family fixed effects, the incremental (within) $R^2$ for the polygenic score is reduced to 1%. This reduction in predictive power is well-established in the literature (see e.g., Koellinger & Harden, 2018; Kong et al., 2018; Lee et al., 2018; Rietveld et al., 2013; Selzam et al., 2019) and reflects the fact that family fixed effects account for the shared family environment and parental genotype, which is not accounted for in the between-family analysis.

*Figure 1. The standardized polygenic score and years of education.*

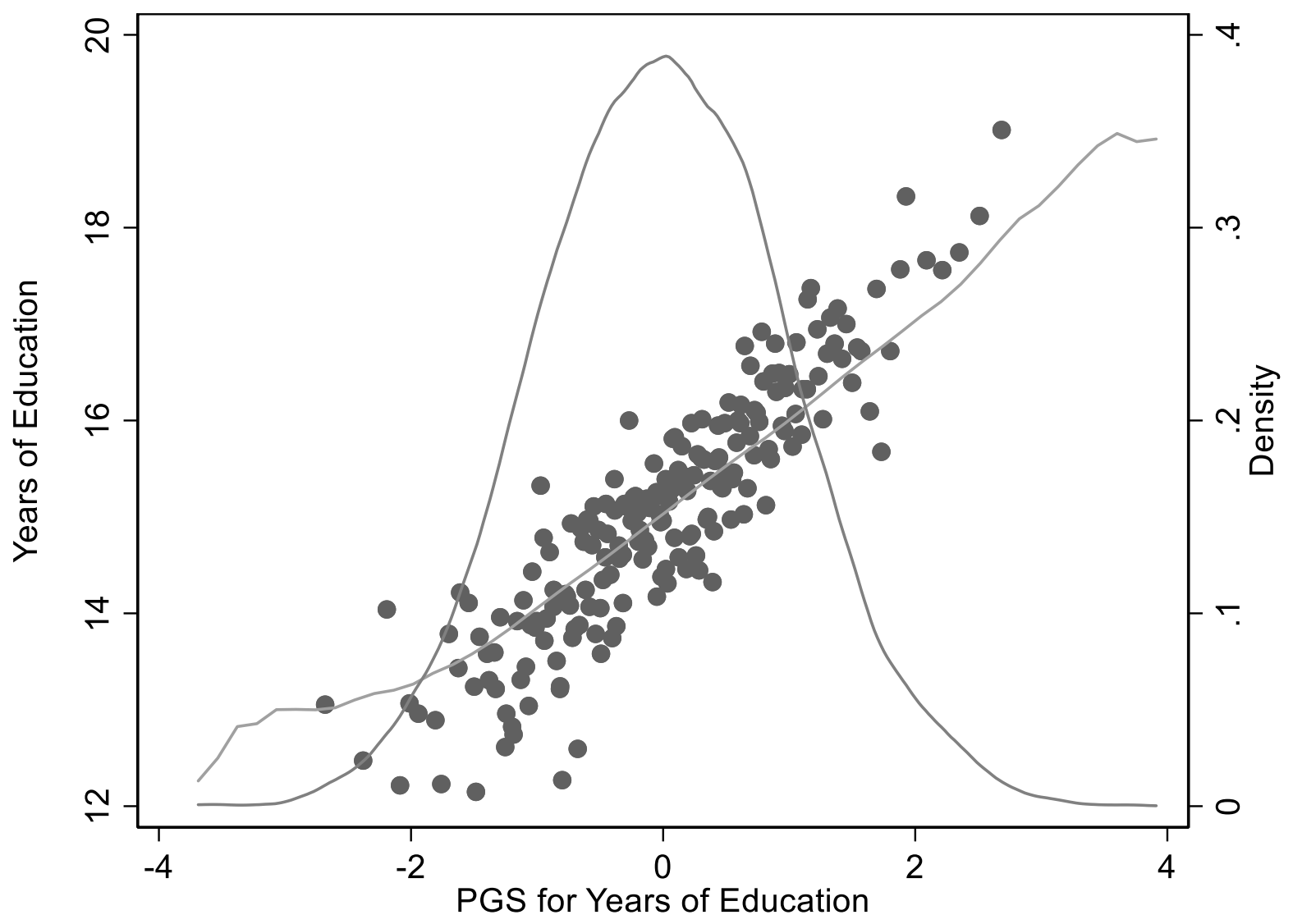

*Notes:* Plot based on 200 bins of the polygenic score distribution. The dots show the average years of education for each bin.

## 4.2. Birth order and educational attainment

Figure 2 shows differences in years of education (residualized with respect to year and month of birth, family size, gender, and the first 40 principal components of the genetic relatedness matrix) by birth order (panel A). Not taking into account family fixed effects, we find that years of education for laterborn children is lower (albeit with larger confidence intervals) than that for firstborns. When pooling all laterborns (panel B), the difference between firstborns and laterborns becomes more pronounced.

Table C.2 in Appendix C confirms the birth order effects in the specifications with and without family fixed effects. We observe a consistent gap of 0.3-0.4 years of schooling between first- and laterborn children. The direction and magnitude of the effect is robust to using the binary indicator or the categorical variable for birth order. The within-family effect size for 5$^{th}$ born children does not reach

statistical significance due to the relatively small number of observations (see Table C.2 in Appendix C).

***Figure 2.*** *Birth order and years of education.*

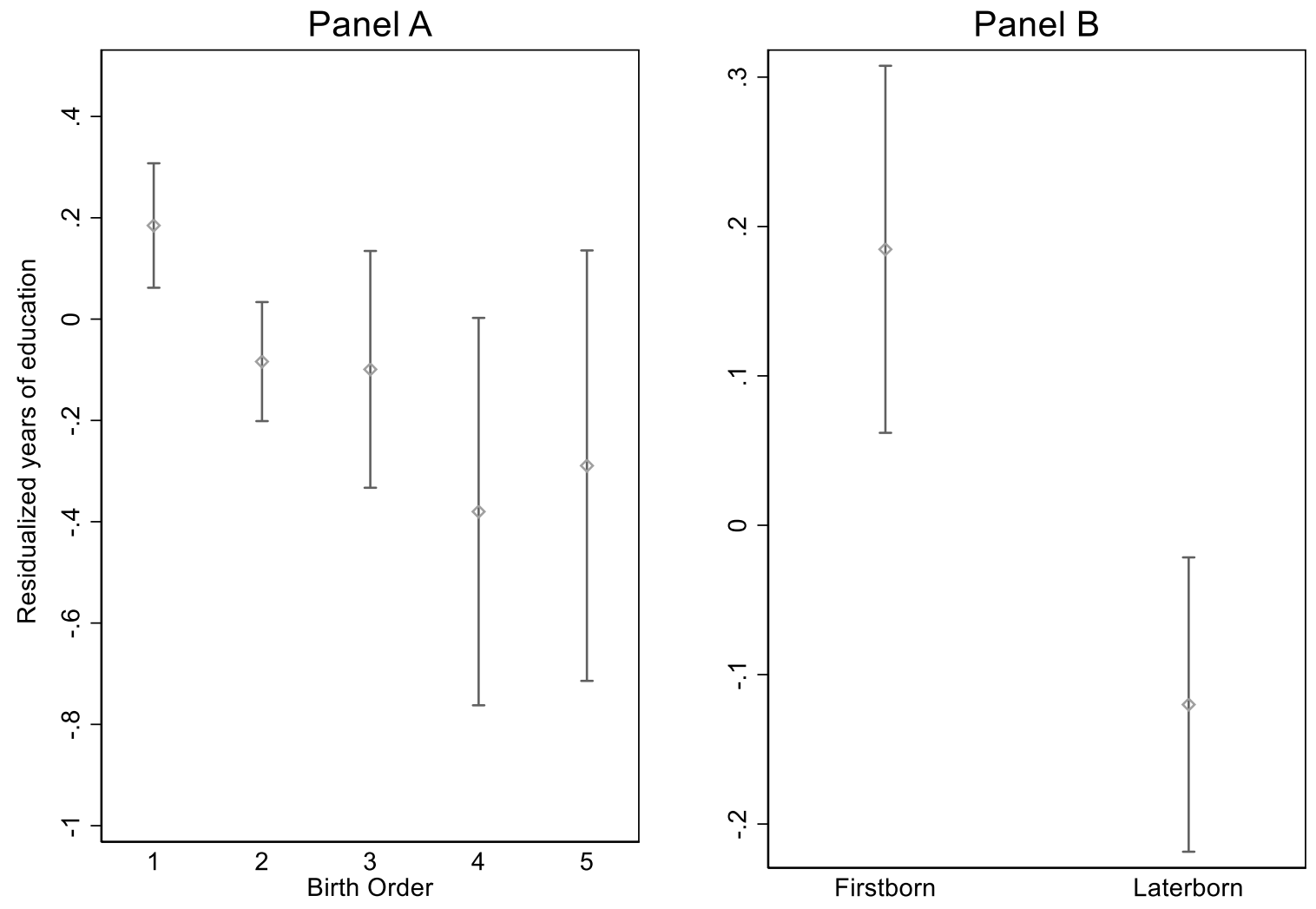

*Notes:* 95% confidence intervals. Years of education is residualized with respect to year and month of birth, family size, gender, and the first 40 principal components of the genetic related matrix.

## 4.3. Birth order and the polygenic score for years of education

To measure causal gene-environment interactions, birth order and the polygenic score need to be orthogonal. Figure 3 provides the first impression of the relationship between the two measures, where the polygenic score for years of education is again residualized using our standard set of controls. Panel A illustrates that the educational attainment polygenic score does not reveal any systematic pattern across birth order. Although Panel A suggests that 3$^{rd}$ born children have somewhat higher genetic endowments, this result is not confirmed in the within-family analysis. The same holds when focusing the comparison on firstborns versus laterborns in Panel B: the observed differences are small and not statistically significant. Furthermore, the distributions of the polygenic score by birth order are overlapping almost perfectly (Panels C and

D).

Consistent with the graphical evidence, Table C.3 in Appendix C shows a slight difference of 0.04 standard deviations in the polygenic score between firstborn and laterborn children without family fixed effects. However, the difference becomes negligible and statistically insignificant within-families. This confirms that firstborns on average do not have different genetic endowments for educational attainment compared to their laterborn siblings (as expected, based on Mendel's Law of Independent Assortment).

***Figure 3.*** *The relationship between birth order and the polygenic score for years of education in the analysis sample.*

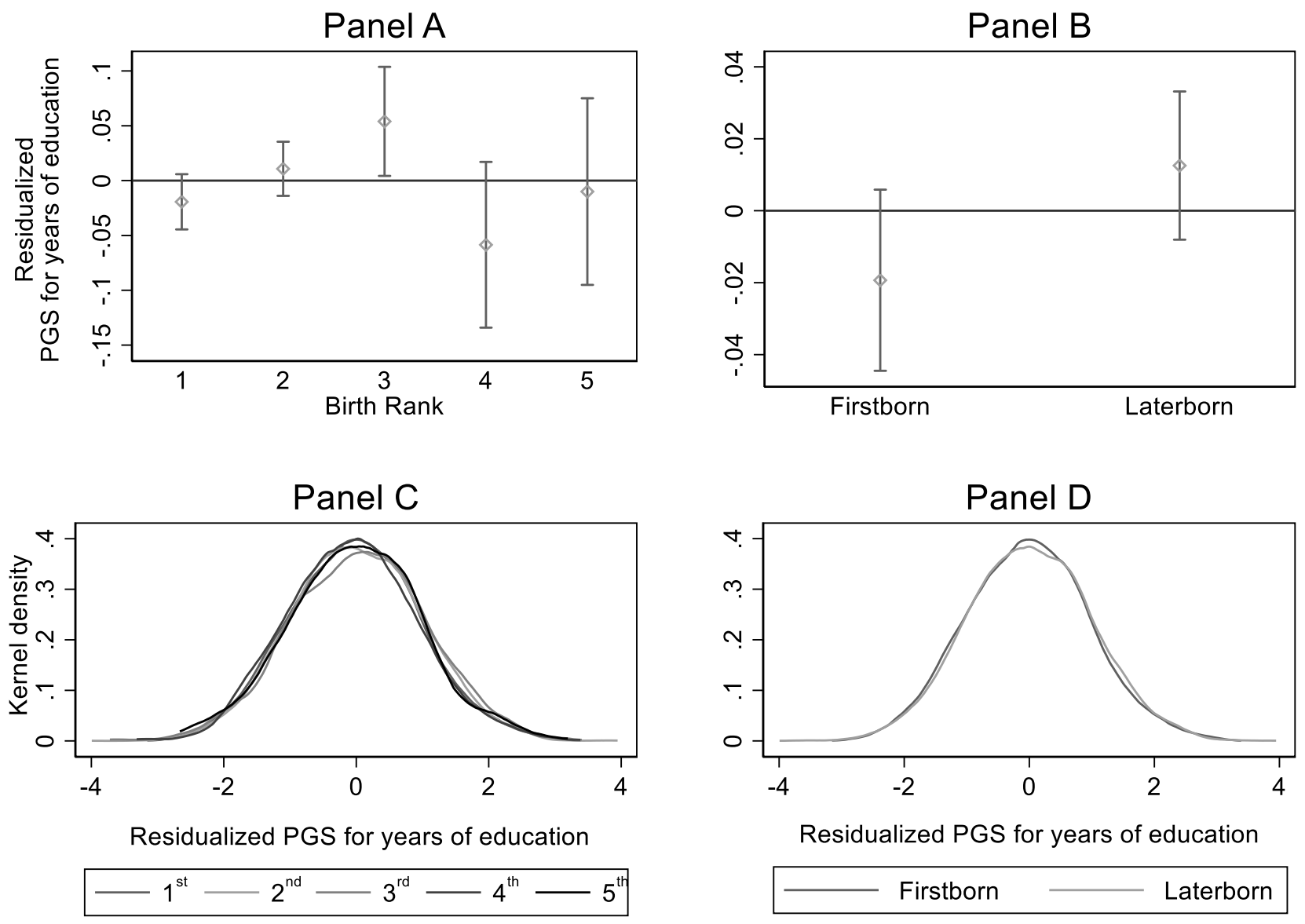

*Notes:* 95% confidence intervals. The polygenic score (PGS) for years of education is residualized with respect to year and month of birth, family size, gender, and 40 first principal components of the genetic related matrix.

This evidence is however not sufficient to claim that birth order is unrelated to *any* genetic endowments. To check if there are systematic differences by birth order in other, possibly related, polygenic scores, we analyze the polygenic scores from the Polygenic Index Repository (Becker, Burik, et al., 2021) for all anthropometric, health, health behavior, and personality traits. We find no systematic difference by birth order in any of the 29 available polygenic scores (see Appendix G for details). These results corroborate the finding that there is no systematic association between birth order and genetic endowments, and hence no evidence for gene-environment correlation (*rGE*).

### 4.4. The gene-environment interaction

Table 2 presents the gene-environment interaction results. Comparison of Columns 1 and 4 in Table 2 to the estimates presented in Tables C.1 and C.2 (Appendix C) shows that including both the educational attainment polygenic score and the indicator for being firstborn does not change their main effects, again indicating independence between the two. Firstborns enjoy on average ~0.4 extra years of schooling compared to laterborns. Furthermore, a one standard deviation increase in the polygenic score is estimated to raise years of education by 1.14 years (without family fixed effects) and 0.65 years (with family fixed effects). For comparison, the effect of parental years of education on offspring years of education is typically estimated to be between 0.2 and 0.3 (Holmlund et al., 2011).

The design without family fixed effects shows a positive interaction between the polygenic score and being firstborn (Column 2), which is borderline statistically significant at conventional thresholds. When we use Obviously-Related Instrumental Variables (ORIV) regression (Column 3), the interaction term becomes larger in magnitude and statistically significant at the 5% threshold, suggesting that measurement error in the polygenic score attenuated the main effect of the polygenic score and the interaction term. In the family fixed effects specification, we find that the interaction effect is significant in both the OLS and the ORIV specifications. The measurement error correction again strengthens the interaction term in terms of magnitude and statistical significance.

*Table 2. Regressions of years of education on the gene-environment interaction.*

| | Without family fixed effects | | | With family fixed effects | | |
|---|---|---|---|---|---|---|
| | (1) OLS | (2) OLS | (3) ORIV | (4) OLS | (5) OLS | (6) ORIV |
| Firstborn | 0.400 | 0.400 | 0.428 | 0.419 | 0.415 | 0.428 |
| | (0.078) | (0.078) | (0.078) | (0.109) | (0.109) | (0.109) |
| PGS for years of education | 1.144 | 1.099 | 1.376 | 0.646 | 0.567 | 0.757 |
| | (0.040) | (0.049) | (0.063) | (0.071) | (0.078) | (0.102) |
| Firstborn × PGS for years of education | | 0.119 | 0.222 | | 0.204 | 0.285 |
| | | (0.072) | (0.097) | | (0.080) | (0.102) |
| Constant | 15.516 | 15.517 | 14.742 | 12.578 | 12.544 | 12.184 |
| | (1.710) | (1.716) | (1.762) | (1.819) | (1.838) | (4.419) |
| $R^2$ | 0.121 | 0.121 | | 0.058 | 0.059 | |
| Cragg-Donald *F*- | | | 3568.570 | | | 2971.412 |
| *N* | 14,850 | 14,850 | 14,850 | 14,850 | 14,850 | 14,850 |

*Notes:* Robust standard errors in parentheses, clustered by family; Coefficients for the control variables (year and month of birth, gender and the first 40 principal components of the genetic relatedness matrix) are not displayed, but available upon request from the authors. We do not report the $R^2$ for the ORIV specifications in Column (3) and (6) given the differences in its interpretation and computation for the instrumental variable type of regressions.

The positive and significant interaction term provides strong evidence for gene-environment interactions in education: those with a higher polygenic score benefit more from being firstborn. The magnitude of the coefficients suggests that for those with a polygenic score two standard deviations below the mean, there is no advantage of being firstborn. In contrast, for those with a high polygenic score, being firstborn increases one's years of education. For example, firstborns with a polygenic score two standard deviations above the mean completed on average 0.82 (OLS) to 1 (ORIV) more years of education compared to their laterborn siblings with a similar polygenic score.[13]

[13] We also investigated potential gender differences in the interaction of interest. This analysis relies on families with a mixed gender composition to ensure variation in gender within families, reducing the analysis sample considerably. While we do find that gender is an

## 4.5. Robustness checks

We next check the robustness of our results against potential non-linearities in the functional form of the polygenic score as well as birth order, against potentially endogenous fertility choices, and against the addition of further control variables. Moreover, we investigate possible issues with identifying an interaction term in a family fixed effects model.

While the linear form seems justified for the polygenic score based on the visual inspection of Figure 1, we explore robustness of our results by allowing for possible non-linearities. Table C.4 in Appendix C compares the within-family specification in continuous form, with those where we specify the polygenic score as binary (above and below the mean), in quartiles and in squared form. We observe positive interaction terms across all specifications. In line with our main findings, the effect of being firstborn is insignificant for those with a lower polygenic score, and the effects are concentrated among those in the top quartile of the polygenic score distribution. We also find that the main result is robust to the inclusion of a quadratic version of the polygenic score.

Table C.5 reports the sensitivity of the results to an alternative specification of birth order. That is, we include dummies for each birth rank with firstborns as the reference category.[14] All point estimates of the main effects and the interaction terms are as expected: on average, secondborns have lower educational attainment compared to firstborns and benefit less from having a high polygenic score. The estimates for higher ranks are rather imprecisely estimated due to the relatively small sample sizes for those with a birth rank of three or higher, but the point estimates consistently suggest that laterborns

---

important predictor of educational attainment for the cohorts examined here, with men having 1.46 more years of education than women, the interaction term does not differ significantly by gender.

[14] Birth rank is constructed from information on the number of older and younger siblings. Hence, this again does not require us to observe all siblings in the UK Biobank.

benefit less from having a high polygenic score compared to firstborns.

In Table C.6 in the Appendix, we explore robustness of our results to potentially endogenous fertility decisions. A possible correlation between our measure of endowments and birth order could arise when fertility decisions are based on the genetic endowments of the children, known in the literature as the "child stopping rule" (Black et al., 2005; Pavan, 2016). Whereas Section 4.3 shows that such a correlation does not exist, here we explicitly control for a possible child-stopping rule by including a dummy variable for being lastborn, which is set to one if an individual's birth order is equal to the total number of children in his/her family. We find that the lastborn dummy is not statistically significant and does not meaningfully affect our results, suggesting that potential endogenous fertility decisions do not change any of our conclusions. To check if our results are sensitive to the exclusion of relatively large families with possibly different characteristics, Table C.6 also includes within-family results based on data from the families with at most three siblings ($N$ = 11,364) and two siblings only ($N$ = 7,918). Even though this sample restriction is endogenous, it is reassuring that the point estimates are virtually identical in these restricted samples.

In Table E.1 in Appendix E we show robustness to the inclusion of additional control variables, including parental age at birth and a set of two-way interactions as suggested by Keller (2014). Specifically, we interact both the dummy for being firstborn and the polygenic score for education with year of birth, month of birth, gender, and first 40 principal components of the genetic relatedness matrix and include all these interactions as additional control variables in the analysis. Our results are robust against these checks.

Finally, we investigate the concern that the interaction term in family fixed effects models may be identified off *between*-family variation (e.g., Shaver, 2019; Giesselmann and Schmidt-Catran, 2022). More specifically, Shaver (2019) shows that Equation 3 can be written as

$$Y_{ij} = \alpha_1 + \alpha_2(G_{ij} - \bar{G}_j) + \alpha_3(E_{ij} - \bar{E}_j) + \alpha_4[(G_{ij} \times E_{ij}) - (\overline{G_{ij} \times E_{ij}})] + \xi_{ij}\,. \tag{5}$$

While the main effects of $G$ and $E$ are identified using only *within*-family variation, the interaction term may be contaminated by unobserved differences *between* families. Although there is no perfect solution to this issue, Shaver (2019) suggests using so-called segmented (or stratified) regressions to ensure that identification of the interaction effect relies on within-family variation only. In Appendix H we show how the effect of the polygenic score is significantly stronger among firstborn siblings than among laterborn siblings, and vice versa, how the effect of being firstborn is significantly stronger for siblings with a polygenic score above the mean compared to those below the mean. These results build confidence that our main results are robust and that our interaction term is primarily identified based on within-family variation.

## V. Discussion

A large literature shows consistently higher educational attainments for firstborn children. Using within-family data we move beyond the existing literature by showing that children benefit disproportionally more from being firstborn when they have higher genetic endowments, as proxied by their polygenic score for educational attainment. More specifically, firstborns with an average polygenic score enjoy 0.428 years (≈ 5 months) of additional schooling compared to their laterborn siblings, on average. However, firstborns with a polygenic score that is two standard deviations above the mean enjoy a full extra year of education, compared to their laterborn siblings with the same genetic endowment. In contrast, for individuals with below-average polygenic scores, being firstborn does not provide an advantage in terms of educational attainment. Since we provide evidence that genetic endowments are orthogonal to birth order, our findings provide one of the first pieces of empirical evidence for the long-held belief that educational attainment is shaped by a complex interplay between genes and environments.

Why would birth order interact with genetic endowments in educational attainment? One possible interpretation of the positive and significant interaction term is the existence of complementarities between endowments and investments in human capital production (e.g. Becker & Tomes, 1986; Cunha & Heckman, 2007). Indeed, the theoretical literature on the 'quantity-quality trade-off' (Becker, 1960; Becker & Lewis, 1973; Becker & Tomes, 1976; Galor & Weil, 2000) posits that with each additional child, it is more expensive to maintain the same 'quality' children (i.e., with the same level of education or health), implying that parents invest less in laterborn children. Moreover, with parental preferences for fairness in investments over equality in outcomes (see e.g., Berry, Dizon-Ross, & Jagnani, 2020), parents distribute their resources equally over their children, leading to natural dilutions in time investments for laterborn children compared to their firstborn siblings (Blake, 1989; Downey, 2001). Indeed, firstborns have undivided attention until the arrival of the second child (Breining et al., 2020), and the empirical literature suggests that a dominant channel through which birth order affects educational attainment is parental time investments (see e.g., Birdsall, 1991; Black, Grönqvist, & Öckert, 2018; Breining et al., 2020; De Haan, 2010; Del Bono et al., 2016; Monfardini & See, 2012; Pavan, 2016; Price, 2008). Hence, while we cannot directly test this mechanism due to data limitations [15], complementarity between parental investments and genetic endowments is consistent with the positive interaction

[15] We do not observe *actual* parental investments and can only proxy for this using birth order. In addition, realized parental investments are endogenous to the child's endowments (Almond & Mazumder, 2013; G. S. Becker & Tomes, 1986; Jere R. Behrman, Pollak, & Taubman, 1982; Breinholt & Conley, 2023; Rosenzweig & Wolpin, 1995; Sanz-De-Galdeano & Terskaya, 2019).

we observe in our data.

An alternative interpretation could be that the additional investments associated with being firstborn are higher for those with higher polygenic scores. That is, the positive interaction effect could also be explained if parents invest more in the firstborn when the firstborn child has a higher polygenic score. While we cannot fully rule out this explanation, we believe this explanation is less plausible for two reasons. First, Breinholt & Conley (2023) and Houmark et al. (2020) show that parenting during infancy is not driven by genetic make-up because these endowments are not clearly expressed yet, and parental investment responses to polygenic scores do not arise before age six. This is long after the typical arrival of subsequent children, and so the most precious time of undivided attention for the firstborn is unlikely to be influenced by – at that time – unobserved differences in polygenic scores.[16] Second, recent evidence suggests that parents compensate for, rather than reinforce, genetic differences in education between siblings. More specifically, Fletcher et al. (2020) show that the association between the education polygenic score and educational attainment is stronger for siblings with the lower polygenic score. This is consistent with parental preferences for equality among siblings, and compensating parental investments, and does not support the alternative interpretation that parents invest more in the firstborn only when the firstborn has a higher polygenic score.

Another potential explanation for our findings is that other mechanisms through which birth order effects arise (e.g., parental age at birth, interactions with younger siblings) could interact with genetic endowments. For example, a recent study suggests that older siblings may influence the education of younger siblings, although it is acknowledged that the estimated relationship may also reflect unaccounted for differences between families (Howe et al., 2022). Accounting for parental age at birth and other potential confounders does not affect our results (see Appendix E). Unfortunately, we cannot test alternative explanations regarding sibling interactions directly because the UK Biobank is very limited in measures of actual parental and sibling interactions. Still, Table C.5 (Appendix C) suggests that our findings are mainly driven by a distinction between first- and all laterborns, with significant differences appearing prominently between the first and secondborns. If interactions with younger siblings would be driving our results, one would expect a more gradual decrease in the magnitude of the interaction terms with rising birth rank, as some second- and thirdborns similarly benefit from interactions with their younger siblings.

[16] For the few early-life parental investments we observe in our data, we do not find evidence of any response to the polygenic score. Appendix D shows that maternal smoking around pregnancy and whether the child was breastfed are all unrelated to the firstborn's polygenic score. If anything, the age gap between first- and secondborns is slightly lower if the firstborn has a higher polygenic score.

A number of limitations should be acknowledged. First, we do not observe parental investments. Future studies should employ richer measures to break down the complementarity between genes and advantageous environments into e.g., parental investments and sibling interactions. A second limitation is that our measure of genetic endowments is imperfect. In particular, a polygenic score captures only common genetic variations in the human genome, and even within the realm of common variations the measure is subject to measurement error. While the use of ORIV reduces concerns about classical measurement error, our family fixed effects estimates of the polygenic score are still subject to attenuation bias due to genetic nurture, such that our effect sizes represent conservative estimates.

The polygenic score should also not be interpreted narrowly as a measure of immutable biological endowments: while within-family analyses allow us to interpret the effect of the polygenic score as a causal effect of genetic variation, it is well-established that the environment may mediate this effect (e.g.,

Breinholt & Conley, 2023; Houmark, Ronda, & Rosholm, 2020), including the family environment (Fletcher et al., 2020). Thus, a polygenic score measures education-enhancing endowments, and will reflect how *on average* in the discovery sample environments respond to differences in genetic endowments. Importantly though, since the measure is fixed at conception and orthogonal to birth order in the within-family analysis, the measure does not reflect investments of the child's parents. As a result, the inclusion of environmental responses to genetic variation into the construction of the polygenic score is not a source of concern for our identification strategy but does imply that our findings may be specific to the context studied.

A related limitation regards the external validity of the empirical findings. As mentioned in Section 3, there is sample selection into the UK Biobank, with a bias towards healthier and higher-educated individuals (Fry et al., 2017). On top of this, we focus on European-ancestry individuals and the coincidental sampling of siblings even though these were not specifically targeted, further reducing the representativeness of the sample. Finally, we construct our polygenic score on basis of a tailor-made GWAS, again on basis of the same UK Biobank excluding the siblings and their relatives. While the latter choice helps to maintain the same environments across discovery and prediction sample, it may further increase the likelihood that our results are specific to the UK Biobank. Therefore, we call for future studies to replicate our findings in other contexts.

# VII. ONLINE APPENDIX to Muslimova et al. (2023)

## A. Genetic data, GWAS, and polygenic scores

*Genetic Data.* A complete human genome consists of 23 pairs of chromosomes, from which the 23rd pair determines the biological sex of a person. One of each pair of chromosomes is inherited from the father, and the other is inherited from the mother. A chromosome is composed of two intertwined strands of deoxyribonucleic acid (DNA), each made up of a sequence of four possible nucleotide molecules: adenine, cytosine, thymine, and guanine. Adenine (A) on one strand is always paired with thymine (T) on the other strand, and cytosine (C) is always paired with guanine (G). These pairs are called base pairs. Every human genome consists of approximately 3 billion base pairs and stretches of base pairs coding for proteins are called genes. There are approximately 20,000 genes in the human genome, with varying lengths in terms of base pairs (Ezkurdia et al., 2014).

Two unrelated human beings share approximately 99.6% of their DNA, and most genetic differences across humans can be attributed to single nucleotide polymorphisms (SNP) (Auton et al., 2015). A SNP is a locus in the DNA at which two different nucleotides can be observed in the population. Each of the two possible nucleotides is called an allele for that SNP. An individual's genotype is coded as 0, 1, or 2, depending on the number of "effect" alleles present. In the human genome, there are at least 85 million SNPs with a "minor" allele prevalence of at least 1% (Auton et al., 2015).

Genome-Wide Association Studies (GWASs) aim to identify genetic variants that are associated with a particular trait of interest by relating each variant to the trait in a hypothesis-free approach. Stringent significance thresholds are used to identify variants that are robustly associated with the trait, with other independent samples used for replication. Using the GWAS approach, thousands of genetic discoveries have been made (Visscher et al., 2017).

Individual SNPs typically explain less than 0.02% of the variance in a behavioral outcome (Chabris, Lee, Cesarini, Benjamin, & Laibson, 2015; Visscher et al., 2017). It is therefore common to combine multiple SNPs into a polygenic score (Dudbridge, 2013), constructed as a weighted sum of SNPs. Through increases in GWAS sample sizes, the predictive power of the polygenic score for education has increased from 2-3% (Rietveld et al., 2013), to 6-8% (Okbay et al., 2016), to 11-13% (Lee et al., 2018), to 12-16% (Okbay et al., 2022). In terms of biological pathways, there is evidence that many of the identified genes associate with health, cognitive, and central nervous system traits (Rietveld et al., 2013). Likewise, the majority of the significant SNPs in Okbay et al. (2016) and Lee et al. (2018) relate to genomic regions responsible for gene expression in a child's brain during the prenatal period.

Methods. *Relatedness.* As a first step, we identify siblings and their relatives using the kinship matrix provided by the UK Biobank. The kinship matrix is based on genetically identified relatedness and contains relatives of third degree and closer identified using the KING software (Manichaikul et al., 2010). The UK Biobank does not have information about self-reported relatedness (Bycroft et al., 2018). The degree of relatedness between the pairs of individuals is based on the combination of the kinship coefficient and genetic similarity in terms of the identity by state ($IBS_0$) coefficient. $IBS_0$ measures the fraction of markers for which the related individuals do not share alleles. We follow the KING manual regarding the thresholds for how to determine family relationship (see Table A.1). The identified number of pairs per relationship type differs slightly from that of Bycroft et al. (2017), because some UK Biobank participants withdrew their consent to analyse their data since then.

For our analyses, we go one step further by separating those who are related to the siblings up to the 3rd degree (kinship coefficient ≥ 0.025), i.e., siblings, parents of siblings, cousins of siblings (See Table A.2). In this way, our holdout sample for polygenic score construction and prediction (i.e., the sibling subsample) is unrelated to the GWAS discovery sample which is used to calibrate the SNP weights that are used to construct the polygenic score.

***Table A.1.*** *Thresholds used to determine relatedness between individuals in*

| | Duplicate / Monozygotic twins | 1st degree / Parent-child | 1st degree siblings | 2nd-3rd degree relatives / cousins | Total |
|---|---|---|---|---|---|
| Kinship coefficient | >0.3540 | 0.1770–0.3540 | 0.1770–0.3540 | 0.0442–0.1770 | |
| $IBS_0$ | | <0.0012 | >0.0012 | | |
| *N* (pairs) | 179 | 6,271 | 22,659 | 78,038 | 107,147 |

***Table A.2.*** *Relatedness to the individuals in the siblings' subsample of UK*

| Relationship to siblings | Unrelated to siblings | Full siblings | 2nd-3rd relative of siblings | Parent or child of siblings | Total |
|---|---|---|---|---|---|
| *N* (individuals) | 91,055 | 41,498 | 10,207 | 4,740 | 147,500 |

*Notes:* Relatedness to siblings is computed based on the relatedness classification as reported in Table A.1.

*GWAS.* Our tailor-made GWAS is performed using the fastGWA protocol for Genome-wide Complex Trait Analysis (GCTA) developed by Jiang et al. (2019). fastGWA applies mixed linear modelling (MLM) to the genetic data of the UK Biobank. fastGWA requires the following steps. First, we generate a sparse genetic relatedness matrix (GRM) using the family relatedness file from the UK Biobank based on the KING software output. Next, we perform an MLM-based GWAS using the SNP data, the sparse GRM, the phenotype file and the minor allele frequency (MAF) filter of 0.001. The phenotype file provides the data on individual years of education residualized with respect to birth year, gender, interaction of birth year and gender, batch, and the first 40 principal components (PCs) of the genetic relatedness matrix. For quality control reasons, some individuals were not included in the GWAS.[17] The eventual GWAS discovery sample includes 389,419 individuals.

We further quality control the resulting GWAS summary statistics using EasyQC tool (Winkler et al., 2014) and meta-analyse our tailor-made GWAS weights with the summary statistics from Okbay et al. (2016). We use these for constructing an alternative polygenic score that is used in the robustness analysis (see footnote 18). Meta-analysis is conducted using the software package METAL (Willer, Li, & Abecasis, 2010).

*Polygenic scores.*

The polygenic score measures the genetic predisposition towards educational attainment within the environmental context and demographic characteristics of the discovery GWAS sample (Domingue, Trejo, Armstrong-Carter, & Tucker-Drob, 2020). It is therefore preferable to select discovery and analysis samples from the same environmental context, especially when analyzing gene-environment interactions. At the same time, the discovery sample should be independent of the analysis sample to avoid overfitting (Dudbridge, 2013). Navigating this trade-off, we therefore construct the polygenic score by using the weights from our own tailor-made GWAS that uses the UK Biobank sample, but without the siblings (our analysis sample) and their relatives. The GWAS discovery sample comprises 389,419 individuals from the UK Biobank; we use the summary statistics from these analyses to create the polygenic scores on the sample of 14,850 siblings. This tailor-made polygenic score alleviates the differences between the discovery and the analysis samples in terms of demographics and environmental context, as well as measurement, i.e., the variables of interest are measured in the same way (Elam, Clifford, Shaw, Wilson, & Lemery-Chalfant, 2019; Tropf et al., 2017). Moreover, running our own GWAS enables the construction of two independent polygenic scores, obtained by splitting the discovery GWAS sample into two equal halves, which can be used in ORIV. This approach has been shown to outperform a single polygenic score that is based on meta-analyzing multiple cohorts (Van Kippersluis et al., 2022).[18]

[17] More specifically, we exclude individuals who withdrew consent, have missing gender or whose self-reported gender does not match the genetic sex, are of other than European ancestry, have bad genotyping quality, putative sex chromosome aneuploidy, whose second chromosome karyotypes are different from XX or XY, with outliers in heterozygosity, or have missing information on any of the former criteria.

The polygenic scores are constructed while accounting for linkage disequilibrium between SNPs using LDpred (Vilhjálmsson et al., 2015), version 1.06, and Python, version 3.6.6. Linkage disequilibrium pertains to the non-random correlations between SNPs at various loci of a single chromosome. LDpred is a software package based on Python that adjusts the GWAS weights for LD using a Bayesian approach. We follow the steps as outlined in Mills, Barban, & Tropf (2020), including the coordination of the base and target files, computing the LD adjusted weights, and then applying them for polygenic score construction using PLINK (Purcell et al., 2007). We re-weight the SNP effects on the basis of LD and the supposed fraction of causal SNPs, which we set to 1, as is standard practice for behavioral traits (Cesarini & Visscher, 2017). Our hold-out sample for constructing polygenic scores consists of 49,866 siblings and their relatives, where the final analysis sample with observations for all variables available is 14,850 individual siblings. The polygenic scores include all SNPs, that is 1,065,078 SNPs after filtering for HapMap3 SNPs at the coordination step. For the split sample GWAS, we first remove all remaining parent-child pairs (*N* = 5,134) and cousins except one from each cousin cluster (*N* = 45,099) and split the unrelated discovery sample with all control variables available (*N* = 340,009) randomly into two samples of 170,005 and 170,004 individuals each and use the same fastGWA procedure as for the full UKB GWAS to obtain SNP weights. The removal of parent-child pairs and cousins ensures that two subsamples do not contain related individuals and are thus independent from each other. We proceed by using LDpred to construct two polygenic scores based on the two sets of summary statistics. Likewise, we include all SNPs (1,065,146 after filtering for HapMap SNPs at the coordination step).

[18] To check if our results are sensitive to the choice of polygenic scores, we constructed a polygenic score based on the meta-analyzed GWAS results of 23andMe summary statistics and our own UK Biobank discovery sample GWAS. As expected, this polygenic score is more predictive for educational attainment than the polygenic score constructed on the basis of the UK Biobank alone. However, this polygenic score is based on several discovery cohorts from very different environmental contexts and does not allow us to use ORIV since we do not have access to all underlying samples to allow us to create multiple polygenic scores. The results are very similar, with the interaction term being slightly smaller but not significantly different from our main results (see Appendix F).

## B. Obviously-Related Instrumental Variable (ORIV) regression

In this section, we describe Obviously-Related Instrumental Variable (ORIV; Gillen et al., 2019) regression. Suppose we would like to predict an outcome variable of interest, *Y*, using a polygenic score, i.e., estimate the following model:

$$Y = \alpha + \beta PGS^{*} + \varepsilon, \tag{B.1}$$

where $\alpha$ is a constant, $\beta$ is the effect of a true polygenic score $PGS^{*}$ and $\varepsilon$ is the error term. We have two estimates of the true polygenic score: $PGS_1 = PGS^{*} + \vartheta_1$ and $PGS_2 = PGS^{*} + \vartheta_2$. The covariance between the two measurement error terms $\vartheta_1, \vartheta_2$ is zero, $Cov(\vartheta_1, \vartheta_2) = 0$, and they have the same relative variance of the measurement errors $\vartheta_1, \vartheta_2$. That is:

$$\frac{\sigma^2_{\vartheta_1}}{\sigma^2_{PGS_1}} = \frac{\sigma^2_{\vartheta_2}}{\sigma^2_{PGS_2}} = \frac{\sigma^2_{\vartheta}}{\sigma^2_{PGS}}, \tag{B.2}$$

where $\sigma^2_{\vartheta_1}$ and $\sigma^2_{\vartheta_2}$ are the variances of the measurement errors $\vartheta_1, \vartheta_2$ respectively, and $\sigma^2_{PGS_1}$ and $\sigma^2_{PGS_2}$ are the variances of respective polygenic scores. If we use $PGS_2$ as an instrumental variable for $PGS_1$, the following applies:

$$plim\ \beta_{IV} = \frac{Cov(Y, PGS_2)/V(PGS_2)}{Cov(PGS_1, PGS_2)/V(PGS_2)} = \frac{Cov(\alpha + \beta PGS^{*} + \varepsilon, PGS^{*} + \vartheta_2)}{Cov(PGS^{*} + \vartheta_1, PGS^{*} + \vartheta_2)} = \beta \frac{\sigma^2_{PGS^{*}}}{\sigma^2_{PGS^{*}}} = \beta. \tag{B.3}$$

ORIV regression as developed by Gillen et al. (2019) estimates a 'stacked' model:

$$\binom{Y}{Y} = \binom{\alpha_1}{\alpha_2} + \beta \binom{PGS_{1+}}{PGS_{2+}} + \varepsilon, \tag{B.4}$$

where one instruments the stack of estimated polygenic scores $\binom{PGS_{1+}}{PGS_{2+}}$ with $\begin{pmatrix} PGS_{2+} & 0_N \\ 0_N & PGS_{1+} \end{pmatrix}$, where *N* is the sample size and $0_N$ is a *N*x1 vector with zero's. We include a family-stack fixed effect to conduct the within-family comparisons within a stack of the data. Standard errors are clustered at both the family and individual level following Correia (2017, 2019).

## C. Supplementary tables

Here, we present tables with supplementary results that are discussed in the main text.

***Table C.1*** *Results of the regressions of years of education on the polygenic score (PGS).*

| | Without family fixed effects | | With family fixed effects | |
|---|---|---|---|---|
| | (1) | (2) | (3) | (4) |
| PGS for years of education | | 1.155 | | 0.646 |
| | | (0.038) | | (0.071) |
| Constant | 15.889 | 15.249 | 13.757 | 13.628 |
| | (1.690) | (1.666) | (1.685) | (1.821) |
| $R^2$ | 0.059 | 0.113 | 0.046 | 0.056 |
| *N* | 14,850 | 14,850 | 14,850 | 14,850 |

*Notes:* Robust standard errors in parentheses, clustered by family; Coefficients for the control variables (year and month of birth, gender and the first 40 principal components of the genetic relatedness matrix) are not displayed, but available upon request from the authors.

***Table C.2*** *Regressions of years of education on different specifications of birth order.*

| | Without family fixed effects | | With family fixed effects | |
|---|---|---|---|---|
| | (1) | (2) | (3) | (4) |
| Firstborn | 0.357 | | 0.418 | |
| | (0.087) | | (0.109) | |
| $2^{nd}$ born | | -0.310 | | -0.450 |
| | | (0.090) | | (0.124) |
| $3^{rd}$ born | | -0.431 | | -0.743 |
| | | (0.146) | | (0.245) |
| $4^{th}$ born | | -0.822 | | -0.808 |
| | | (0.228) | | (0.366) |
| $5^{th}$ born | | -0.892 | | -0.463 |
| | | (0.279) | | (0.498) |
| Constant | 16.258 | 16.473 | 12.711 | 13.002 |
| | (1.746) | (1.737) | (1.681) | (1.678) |
| $R^2$ | 0.069 | 0.069 | 0.048 | 0.048 |
| *N* | 14,850 | 14,850 | 14,850 | 14,850 |

*Notes:* Robust standard errors in parentheses, clustered by; Coefficients for the control variables (year and month of birth, gender and the first 40 principal components of the genetic relatedness matrix) are not displayed, but available upon request from the authors.

***Table C.3** Regressions of polygenic score for educational attainment on birth order.*

| | Without family fixed effects | | With family fixed effects | |
|---|---|---|---|---|
| | (1) | (2) | (3) | (4) |
| Firstborn | -0.037 | | -0.002 | |
| | (0.018) | | (0.018) | |
| 2nd born | | 0.034 | | 0.008 |
| | | (0.019) | | (0.020) |
| 3rd born | | 0.078 | | 0.009 |
| | | (0.031) | | (0.039) |
| 4th born | | -0.036 | | 0.018 |
| | | (0.045) | | (0.057) |
| 5th born | | -0.013 | | 0.061 |
| | | (0.057) | | (0.080) |
| Constant | 0.648 | 0.605 | 0.207 | 0.226 |
| | (0.356) | (0.355) | (0.377) | (0.378) |
| $R^2$ | 0.017 | 0.018 | 0.013 | 0.013 |
| *N* | 14,850 | 14,850 | 14,850 | 14,850 |

*Notes:* Robust standard errors in parentheses, clustered by; Coefficients for the control variables (year and month of birth, gender and the first 40 principal components of the genetic relatedness matrix)) are not displayed, but available upon request from the authors.

*Table C.4 Regressions of years of education on the gene-environment interaction; Robustness to non-linearities in the polygenic score.*

| | With family fixed effects | | | |
|---|---|---|---|---|
| | (1) | (2) | (3) | (4) |
| Firstborn | 0.415 | 0.239 | 0.186 | 0.437 |
| | (0.109) | (0.145) | (0.198) | (0.125) |
| PGS for years of education | 0.567 | | | 0.566 |
| | (0.078) | | | (0.078) |
| Firstborn × PGS for years of education | 0.204 | | | 0.205 |
| | (0.080) | | | (0.081) |
| PGS for years of education (>mean) | | 0.588 | | |
| | | (0.136) | | |
| Firstborn × PGS for years of education (>mean) | | 0.356 | | |
| | | (0.169) | | |
| PGS for years of education (2nd quartile) | | | 0.439 | |
| | | | (0.190) | |
| PGS for years of education (3rd quartile) | | | 0.810 | |
| | | | (0.197) | |
| PGS for years of education (4th quartile) | | | 1.230 | |
| | | | (0.208) | |
| PGS for years of education (2nd quartile) × Firstborn | | | 0.155 | |
| | | | (0.276) | |
| PGS for years of education (3rd quartile) × Firstborn | | | 0.190 | |
| | | | (0.255) | |
| PGS for years of education (4th quartile) × Firstborn | | | 0.595 | |
| | | | (0.232) | |
| PGS for years of education (squared) | | | | 0.018 |
| | | | | (0.047) |
| PGS for years of education (squared) × Firstborn | | | | -0.022 |
| | | | | (0.061) |
| Constant | 12.544 | 12.087 | 11.821 | 12.518 |
| | (1.838) | (1.796) | (1.868) | (1.838) |
| $R^2$ | 0.059 | 0.053 | 0.057 | 0.059 |
| *N* | 14,850 | 14,850 | 14,850 | 14,850 |

*Notes:* Robust standard errors in parentheses, clustered by family; Coefficients for the control variables (year and month of birth, gender and the first 40 principal components of the genetic relatedness matrix) are not displayed, but available upon request from the authors.

*Table C.5 Regressions of years of education; Robustness to non-linearities in birth order.*

| | With family fixed effects | |
|---|---|---|
| | (1) | (2) |
| PGS for years of education | 0.567 | 0.767 |
| | (0.078) | (0.086) |
| Firstborn | 0.415 | |
| | (0.109) | |
| Firstborn × PGS for years of education | 0.204 | |
| | (0.080) | |
| 2nd born | | -0.450 |
| | | (0.124) |
| 3rd born | | -0.741 |
| | | (0.244) |
| 4th born | | -0.832 |
| | | (0.362) |
| 5th born | | -0.470 |
| | | (0.499) |
| 2nd born × PGS for years of education | | -0.212 |
| | | (0.083) |
| 3rd born × PGS for years of education | | -0.152 |
| | | (0.148) |
| 4th born × PGS for years of education | | -0.316 |
| | | (0.254) |
| 5th born × PGS for years of education | | -0.089 |
| | | (0.300) |
| Constant | 12.544 | 12.882 |
| | (1.838) | (1.842) |
| $R^2$ | 0.059 | 0.060 |
| *N* | 14,850 | 14,850 |

*Notes:* Robust standard errors in parentheses, clustered by family; Coefficients for the control variables (year and month of birth, gender and the first 40 principal components of the genetic relatedness matrix) are not displayed, but available upon request.

***Table C.6*** *Regressions of years of education on the gene-environment interaction; Robustness to fertility choices.*

| | With family fixed effects | | | |
|---|---|---|---|---|
| | (1) | (2) | (3) | (4) |
| Firstborn | 0.415 | 0.392 | 0.439 | 0.466 |
| | (0.109) | (0.128) | (0.132) | (0.178) |
| PGS for years of education | 0.567 | 0.568 | 0.589 | 0.570 |
| | (0.078) | (0.078) | (0.090) | (0.109) |
| Firstborn × PGS for years of education | 0.204 | 0.203 | 0.197 | 0.183 |
| | (0.080) | (0.080) | (0.087) | (0.102) |
| Lastborn | | -0.049 | | |
| | | (0.136) | | |
| Male | 1.265 | 1.265 | 1.429 | 1.571 |
| | (0.099) | (0.099) | (0.109) | (0.130) |
| Constant | 12.544 | 12.509 | 13.065 | 12.578 |
| | (1.838) | (1.840) | (2.106) | (2.148) |
| $R^2$ | 0.059 | 0.059 | 0.066 | 0.073 |
| *N* | 14,850 | 14,850 | 11,364 | 7,918 |

*Notes:* Column 1 replicates our main results from Table 2; Column 2 additionally controls for a dummy indicating whether the individual is the lastborn; Column 3 restricts the sample to families with less than four siblings; Columns 4 restricts the sample to families with two siblings only. Robust standard errors in parentheses, clustered by family; Coefficients for the control variables (year and month of birth, family size, and the first 40 principal components of the genetic relatedness matrix) are not displayed, but available upon request from the authors.

## D. Early-life parental investments

The only early-life parental investments observed in the UK Biobank are whether the child was breastfed and whether the mother smoked around birth. We also observe the age gap between subsequent siblings. Table D.1 reports the results of regressions explaining the few early-life parental investment as a function of being firstborn, the polygenic score for education, and their interaction. This shows whether mothers change their behavior depending on whether the child is first- or laterborn and the polygenic score of their children. The results show that the probability of being breastfed (Column 1) and the likelihood of maternal smoking around pregnancy (Column 2) are similar between first- and laterborns. Furthermore, the coefficient of the polygenic score for (the child's) education is not significantly different from zero, and we find no evidence of any differences in maternal investments around pregnancy by the firstborn's polygenic score.

Next, we explore the relationship between the polygenic score and the age gap between siblings. Column 3 presents the estimates from a regression of the age gap in months between every two consecutive siblings on the polygenic score of the older sibling in the pair. These between-family estimates suggest that birth spacing is one month shorter for every standard deviation increase in the polygenic score of the older sibling. However, note that these results derive from a between-family comparison, and therefore should not be interpreted causally. For families for whom we observe three consecutive siblings or more ($N$ = 2,542), we can check this result in a within-family specification. The within-family estimates (Column 4) are very noisy because of the small sample size, and the wide confidence intervals do not allow drawing any firm conclusions here.

***Table D.1*** *Regressions of early life parental investments on the gene-environment interaction.*

| | Breastfed | Mother smoked around birth | Age gap | |
|---|---|---|---|---|
| | With family fixed effects | | Without family fixed effects | With family fixed effects |
| | (1) | (2) | (3) | (4) |
| Firstborn | 0.006 | 0.006 | | |
| | (0.010) | (0.006) | | |
| PGS for years of education | -0.001 | 0.001 | | |
| | (0.007) | (0.004) | | |
| Firstborn × PGS for years of education | -0.011 | 0.004 | | |
| | (0.007) | (0.004) | | |
| PGS for years of education of the older sibling | | | -0.970 | 6.417 |
| | | | (0.331) | (3.466) |
| Constant | 0.705 | 0.270 | 50.444 | -24.447 |
| | (0.067) | (0.040) | (6.934) | (31.458) |
| $R^2$ | 0.042 | 0.020 | 0.067 | 0.376 |
| *N* | 11,818 | 13,156 | 6,501 | 2,542 |

*Notes:* Robust standard errors in parentheses, clustered by family; Coefficients for the control variables (year and month of birth, gender, family size, and the first 40 principal components of the genetic relatedness matrix) are not displayed, but available upon request from the authors. Sample sizes vary depending on the availability of early life parental investments and the number of siblings included in the analyses.

## E. Additional confounders

In Table E.1 we present further analyses to investigate robustness of our results against the inclusion of additional confounders. Column 1 reproduces our main results (Table 2, Column 5). In Column 2 and 3, we employ the correction for missing confounders in gene-environment interaction analyses suggested by Keller (2014) without (Column 2) and with (Column 3) corrections for measurement error. Specifically, we interact both the dummy for being firstborn and the polygenic score for education with year of birth, month of birth, gender, and first 40 principal components of the genetic relatedness matrix and include all these interactions as additional control variables in the analysis. The direct effect sizes of being firstborn and the polygenic score for years of education are now relative to the reference categories of the control variables (born in 1937, born in January, being female). The standard error of the interaction term increases due to the much larger number of regressors in both columns. Whereas the magnitude of the interaction term drops somewhat in Column 2, the point estimates in both Columns 2 and 3 do not significantly differ from our baseline specification in Column 1, and show a robust interaction effect, at least after correction for measurement error.

We also explore whether our results are robust against the inclusion of parental age as an additional control, as this may be one mechanism through which birth order effects arise (Eirnæs & Pörtner, 2004). In our sample, we have only 5,601 individuals with information about the age of the mother and 3,085 individuals with information about the age of the father (both measured at time of birth of the child). The inclusion of these variables as control variables in Column 4 and Column 5 respectively of Table E.1. and the subsequent drop in the sample size decreases the statistical power of our tests. Indeed, the interaction term loses statistical significance when controlling for father's age at birth. However, the effect sizes of the interaction term are similar to our main results, and if anything, are larger.

*Table E.1 Results of the regressions of years of education on the gene-environment interaction and additional control variables.*

| | With family fixed effects | | | | |
|---|---|---|---|---|---|
| | (1) | (2) | (3) | (4) | (5) |
| Firstborn | 0.415 | -10.488 | 1.397 | 0.199 | 0.021 |
| | (0.109) | (3.777) | (1.589) | (0.174) | (0.220) |
| PGS for years of education | 0.567 | -4.870 | -3.577 | 0.576 | 0.528 |
| | (0.078) | (1.286) | (1.644) | (0.127) | (0.155) |
| Firstborn × PGS for years of education | 0.204 | 0.150 | 0.229 | 0.265 | 0.235 |
| | (0.080) | (0.092) | (0.132) | (0.125) | (0.160) |
| Mother's age at birth | | | | 0.099 | |
| | | | | (0.088) | |
| Father's age at birth | | | | | -0.028 |
| | | | | | (0.148) |
| Constant | 12.544 | 25.753 | -53.432 | 12.376 | 6.177 |
| | (1.838) | (3.019) | (4.503) | (2.976) | (2.278) |
| $R^2$ | 0.059 | 0.076 | - | 0.062 | 0.093 |
| *N* | 14,850 | 14,850 | 14,850 | 5,601 | 3,085 |

*Notes:* Robust standard errors in parentheses, clustered by family; Coefficients for the control variables (year and month of birth, gender and the first 40 principal components of the genetic relatedness matrix) are not displayed, but available upon request from the authors. Column 1 replicates our main results for comparison. Column 2 includes as additional control variables the interactions between firstborn and the polygenic score with year of birth, month of birth, gender and the first 40 principal components. Column 3 provides the ORIV estimation of the specification in Column 2. Columns 4 and 5 include controls for maternal age at birth and paternal age at birth, respectively.

## F. Replication of main results using a different polygenic score for educational attainment

Columns 1-3 of Table F.1 replicate the main analyses using the polygenic score based on the meta-analysis of the GWAS summary statistics of our own UK Biobank GWAS described in Section 3 and the GWAS summary statistics of 23andMe (Okbay et al., 2016). The procedure for constructing the polygenic score is identical to the one used in the main analysis. From Columns 1-2, we can see that the incremental $R^2$ of the polygenic scores is 1.1% in the within-family specification. The GxE interaction term (Column 3) is slightly smaller than in the main analysis, but not significantly different from our main result.

***Table F.1*** *Regressions of years of education on the gene-environment interaction with the meta-analyzed polygenic scores (PGS).*

| | With family fixed effects | | |
|---|---|---|---|
| | (1) | (2) | (3) |
| | OLS | OLS | OLS |
| PGS for years of education | | 0.664 | 0.616 |
| | | (0.069) | (0.077) |
| Firstborn | | | 0.436 |
| | | | (0.109) |
| Firstborn × PGS for years of education | | | 0.136 |
| | | | (0.079) |
| Constant | 13.757 | 13.408 | 12.348 |
| | (1.685) | (1.823) | (1.838) |
| $R^2$ | 0.046 | 0.057 | 0.060 |
| *N* | 14,850 | 14,850 | 14,850 |

*Notes:* Robust standard errors in parentheses, clustered by family; Coefficients for the control variables (year and month of birth, gender and the first 40 principal components) are not displayed, but available upon request from the authors; The PGS for years of education in Columns 1-3 is based on a meta-analysis of summary statistics from our own UK Biobank GWAS and 23andMe.

## G. Systematic differences in polygenic scores for other traits by birth order

In this section, we test if there are any systematic differences in polygenic scores for anthropometric, health, and personality traits by birth order. Polygenic scores for these analyses are obtained from the Polygenic Index Repository. These polygenic scores are constructed using PLINK 2 (Chang et al., 2015), where SNPs are corrected for linkage disequilibrium using LDpred (Vilhjálmsson et al., 2015; for further technical details, see Becker, Burik, et al., 2021). Table G.1 reports the results of regressions of the single-trait polygenic scores on a dummy for being firstborn and control variables for year and month of birth, gender, the first 40 principal components of the genetic relatedness matrix, and family fixed effects. All associations are very small in magnitude, and only 2 out of 28 individually reach statistical significance at the 5% level. After a Bonferroni correction for multiple testing (i.e., with a Bonferroni-corrected p-value significance threshold of 0.0018 (=0.05/28)), being firstborn is not significantly associated with any of the selected polygenic scores.

***Table G.1*** *Within-family regressions explaining selected single-trait polygenic scores from the Polygenic Index Repository (Becker et al., 2021) on a dummy for being firstborn.*

| Trait | Coefficient | SE | *p*-value |
|---|---|---|---|
| *Anthropometric* | | | |
| 1. Body mass index | 0.002 | (0.019) | 0.903 |
| 2. Height | -0.018 | (0.018) | 0.323 |
| *Health and health behaviors* | | | |
| 3. Alcohol misuse | 0.050 | (0.019) | 0.008 |
| 4. Asthma | -0.021 | (0.019) | 0.263 |
| 5. Asthma/eczema/rhinitis | -0.016 | (0.019) | 0.393 |
| 6. Attention deficit hyperactivity disorder (ADHD) | 0.000 | (0.019) | 0.985 |
| 7. Cannabis use | 0.045 | (0.019) | 0.015 |
| 8. Cigarettes per day | 0.002 | (0.019) | 0.904 |
| 9. Depressive symptoms | 0.004 | (0.019) | 0.822 |
| 10. Drinks per week | 0.026 | (0.019) | 0.165 |
| 11. Ever smoker | 0.026 | (0.019) | 0.163 |
| 12. Hayfever | 0.002 | (0.019) | 0.936 |
| 13. Migraine | 0.012 | (0.019) | 0.520 |
| 14. Nearsightedness | 0.006 | (0.019) | 0.730 |
| 15. Physical activity | 0.039 | (0.019) | 0.957 |
| 16. Self-rated health | 0.001 | (0.019) | 0.044 |
| *Personality and well-being* | | | |
| 17. Adventurousness | 0.001 | (0.019) | 0.959 |
| 18. Extraversion | -0.009 | (0.019) | 0.623 |
| 19. Left out of social activity | -0.014 | (0.019) | 0.478 |
| 20. Life satisfaction, family | -0.026 | (0.019) | 0.162 |
| 21. Life satisfaction, friends | -0.004 | (0.019) | 0.840 |
| 22. Morning person | -0.000 | (0.019) | 0.987 |
| 23. Narcissism | -0.030 | (0.019) | 0.121 |
| 24. Neuroticism | 0.007 | (0.019) | 0.692 |
| 25. Openness | -0.014 | (0.019) | 0.458 |
| 26. Religious attendance | -0.023 | (0.019) | 0.218 |
| 27. Risk tolerance | 0.049 | (0.019) | 0.009 |
| 28. Subjective well-being | -0.020 | (0.019) | 0.273 |

*Notes: N*=14,835; Robust standard errors (SEs) in parentheses; The Bonferroni corrected significance threshold is 0.05/28 = 0.0018. Coefficients for the control variables (year and month of birth, gender and the first 40 principal components) are not displayed, but available upon request from the authors.

## H. Interaction terms in family fixed effects models

Our baseline regression (Equation (3) in the main text, omitting covariates $X_{ij}$ for simplicity) for individual *i* in family *j* is:

$$Y_{ij} = \alpha_1 + \alpha_2 G_{ij} + \alpha_3 E_{ij} + \alpha_4 (G_{ij} \times E_{ij}) + \delta_j + \xi_{ij}. \quad \text{(H.1)}$$

Interpreted in terms of deviations from the family mean, this model is equivalent to (see also Equation (5) and Shaver (2019)):

$$Y_{ij} = \alpha_1 + \alpha_2 (G_{ij} - \overline{G_j}) + \alpha_3 (E_{ij} - \overline{E_j}) + \alpha_4 [(G_{ij} \times E_{ij}) - (\overline{G_{ij} \times E_{ij}})] + \xi_{ij}. \quad \text{(H.2)}$$

While the coefficients $\alpha_2$ and $\alpha_3$ are identified from within-family variation only, this is not necessarily the case for the interaction term $\alpha_4$ (Shaver, 2019). That is, the deviation from the within-family product $(G_{ij} \times E_{ij})$, or $\alpha_4$, is not the same as the product of the within-family deviations in $G_{ij}$ and $E_{ij}$. This could lead to between-family variation partially identifying the interaction coefficient.

There is no perfect solution to this issue. The interaction between the family deviations $(G_{ij} - \overline{G_j})$ and $(E_{ij} - \overline{E_j})$ is not a fixed effect specification as it would be a nonlinear transformation of the original interaction term since some observations would interact two negative values. It therefore generally does not generate meaningful estimates (Shaver, 2019). One option, suggested by Shaver (2019), is to stratify the sample by one of the two variables that are interacted to avoid having to estimate interaction terms ('a segmented regression'). In our setting, this implies (i) estimating a family fixed effects regression of EA on the PGS separately for firstborns and laterborns; and (ii) estimating a family fixed effects regression of EA on firstborn separately for those above and below the (e.g.) mean of the PGS. If there is a true interaction effect within-families, we would expect to see in (i) a larger coefficient of the PGS for firstborns, and in (ii) a larger firstborn effect for those with a value above the mean in the PGS distribution.

One complication with (i) is that we cannot estimate a family fixed effects specification for the sample of firstborns since there is only one firstborn per family. Instead of the family fixed effects specification, we can rely on the 'deviation specification':

$$Y_{ij} = \alpha_1 + \alpha_2 G_{ij} + \alpha_2^* \overline{G_J} + \xi_{ij}, \tag{H.3}$$

with $\overline{G_J}$ being the mean PGS of the family. Equation (H.3) is formally equivalent to a family fixed effects specification but has the advantage that it is possible to be estimated separately for first- and laterborns.

Table H.1 presents the results for Equation (H.3) separately for firstborn and laterborns using OLS (columns 1 and 2) and ORIV (columns 3 and 4). Consistent with the interaction term being identified at least partly based on within-family variation, we see that the coefficient of the PGS is consistently higher among firstborns than among laterborns. The segmented regressions in Columns 1 and 2 were estimated by OLS, and so here we can test whether the coefficients are significantly different from each other using seemingly unrelated regression (SUR). The corresponding $\chi^2$-test suggests that the OLS coefficients in Columns 1 and 2 are not significantly different ($p$ = 0.371).

To test whether the difference in IV estimates across Columns 3 and 4 are significant, we rely on permutation inference as SUR requires exogenous regressors. To do so, we separately run the ORIV regression of Equation (H.3) for firstborns and laterborns and take the difference between the two coefficients (i.e., 1.168-0.679=0.489). Then we randomly permute both the PGS and birth order independently 500 times and compute the empirical $p$-value as the fraction of times the permuted difference across the two ORIV regressions is larger than the actual difference. Figure H.1 shows the resulting distribution of permuted differences, with our actual difference of 0.489 far beyond the empirically computed critical values (i.e., $p < 0.001$). Therefore, we conclude that the coefficients for the PGS for years of education in the ORIV regressions in Columns 3 and 4 are significantly different from each other.

***Table H.2*** *Regressions of years of education on the polygenic score (PGS) with the mean PGS of the family as control variable.*

| | (1) OLS Firstborn | (2) OLS Laterborn | (3) ORIV Firstborn | (4) ORIV Laterborn |
|---|---|---|---|---|
| PGS for years of education | 0.747<br>(0.131) | 0.587<br>(0.100) | 1.168<br>(0.181) | 0.679<br>(0.135) |
| *N* | 5,849 | 9,001 | 5,849 | 9,001 |

*Notes:* Robust standard errors in parentheses, clustered by family; Coefficients for the control variables (family mean of the PGS, year and month of birth, gender and the first 40 principal components) are not displayed, but available upon request from the authors.

***Figure H.1*** *Distribution of the difference in coefficients for the polygenic score in the subsamples of firstborns and laterborns based on permutation.*

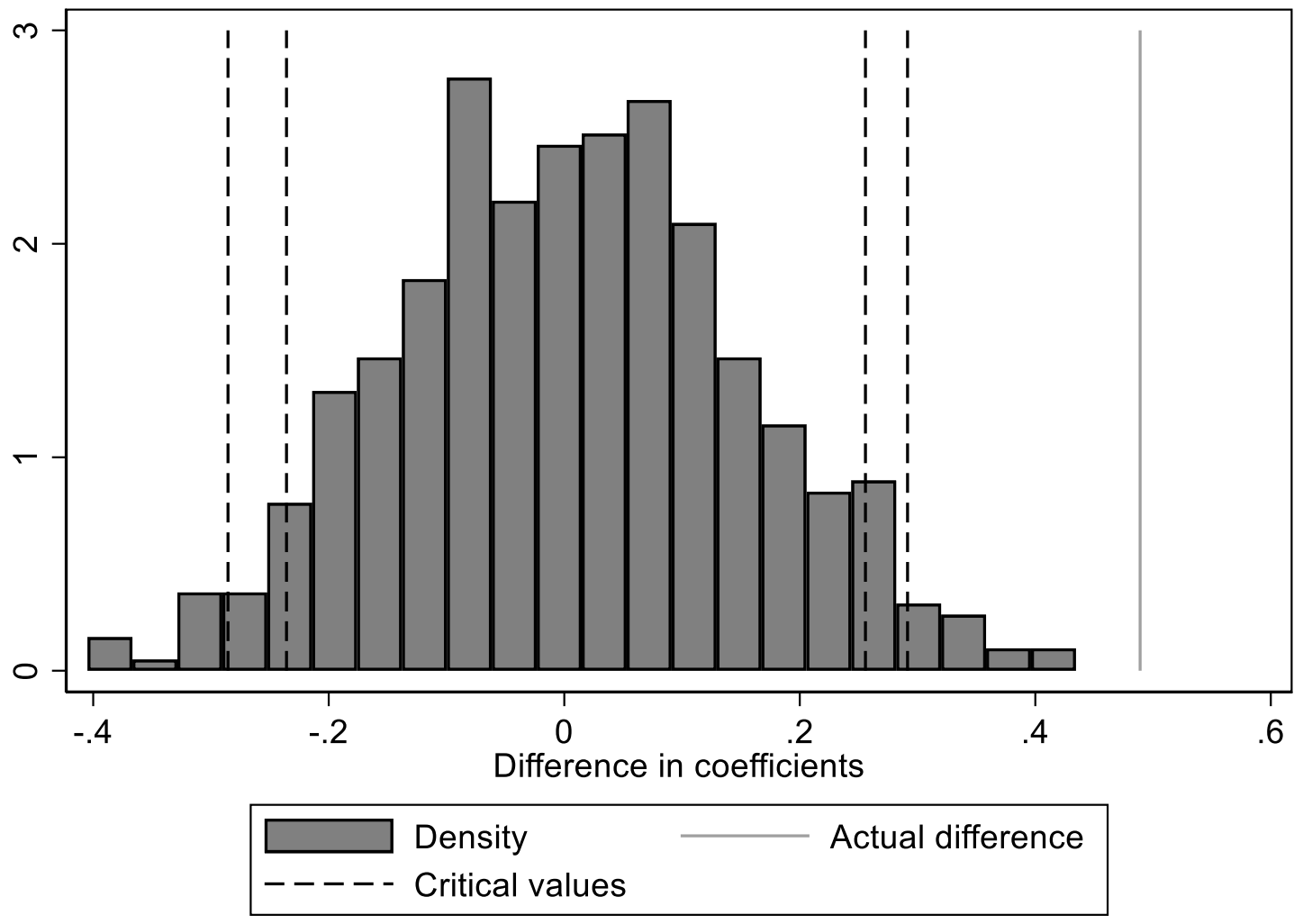

Table H.2 estimates family fixed effects regressions of years of education on the firstborn dummy for those scoring above and below the mean of the distribution of the polygenic score separately. In line with the presence of a true interaction effect within-families, we find a larger firstborn effect for those with a value above the mean in the PGS distribution (0.752 vs. 0.282). Again, using

permutation inference with 500 permutations, Figure H.2 shows the distribution of the difference in coefficients. The actual difference (0.752 – 0.282 = 0.470) lies in between the 95th and 97.5th percentile, and thus the difference in OLS-coefficients is significant at the 5% level for a one-sided test and at the 10% level for a two-sided test.

***Table H.2.*** *Regressions of years of education on firstborn.*

| | With family fixed effects | |
|---|---|---|
| | (1) | (2) |
| | OLS | OLS |
| | Below-mean PGS value | Above-mean PGS value |
| Firstborn | 0.282 | 0.752 |
| | (0.197) | (0.171) |
| *N* | 7,425 | 7,425 |

*Notes:* Robust standard errors in parentheses, clustered by family; Coefficients for the control variables (year and month of birth, gender and the first 40 principal components) are not displayed, but available upon request from the authors.

***Figure H.2.*** *Distribution of the difference in coefficients for firstborn in the subsamples of individuals scoring above and below the mean of the distribution of the polygenic score based on permutation.*

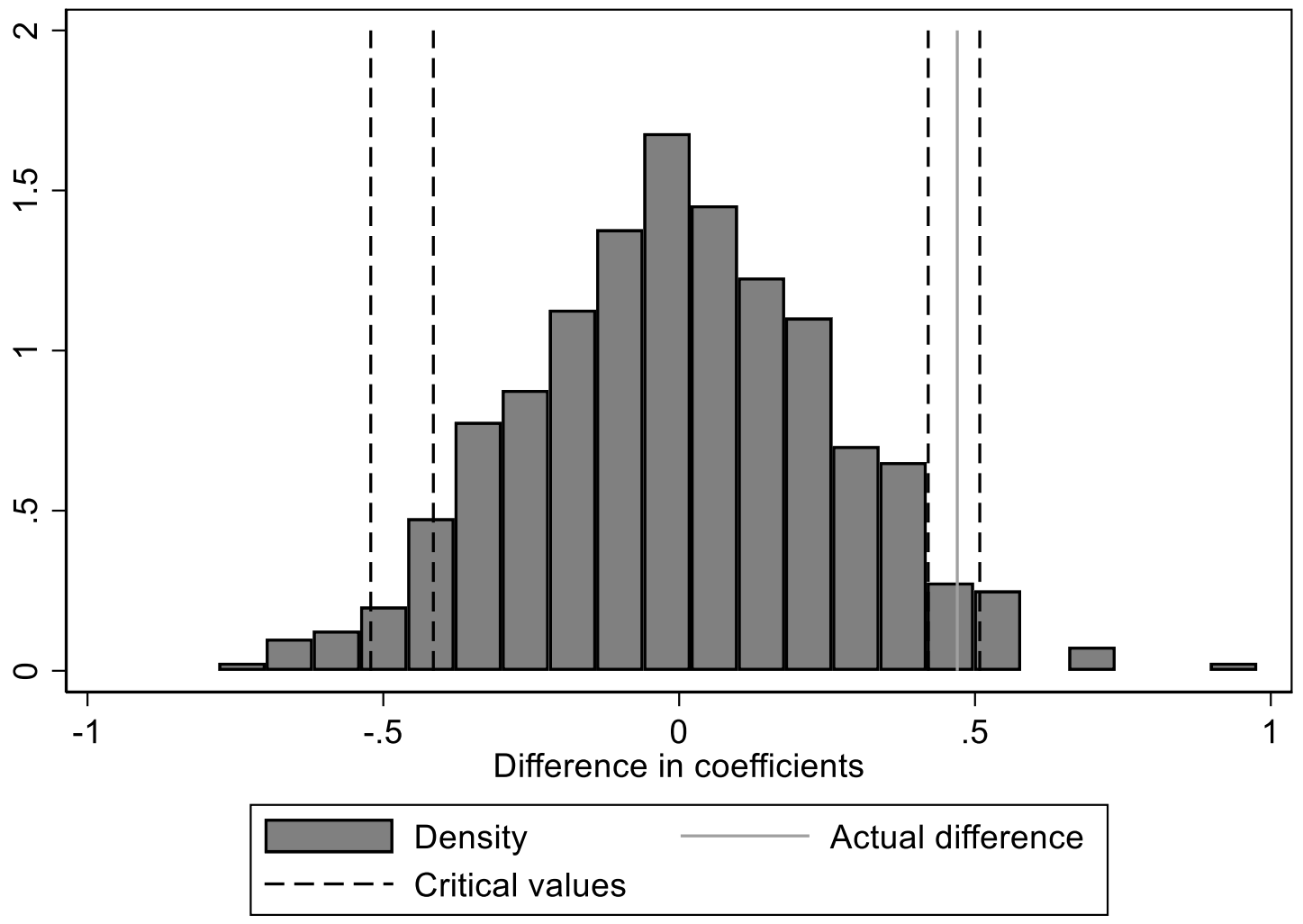